\documentclass{aip-cp}

\usepackage[numbers]{natbib}
\usepackage{rotating}
\usepackage{graphicx}
\newcommand{\hs}{HS\,1946+7658} 
\def\farcs{\hbox{$.\!\!^{\prime\prime}$}}

\begin{document}

\title{A Study of the High-Luminosity Quasar \hs}

\author[]{B.~Mihov\corref{cor1}}
\author[]{L.~Slavcheva-Mihova\corref{cor2}}

\affil{Institute of Astronomy and NAO, Bulgarian Academy of Sciences, 72 Tsarigradsko Chaussee Blvd., BG-1784, Sofia, Bulgaria}
\corresp[cor1]{Corresponding author: bmihov@astro.bas.bg}
\corresp[cor2]{lslav@astro.bas.bg}

\maketitle

\begin{abstract}
We study the variability of the quasar \hs\ on intra-night time scale based on both our own optical and archival X-ray data. We find the quasar non-variable during about 11 hours of optical monitoring. This is in accordance with the low intra-night variability duty cycle of radio-quiet quasars. Regarding the X-rays, we cannot make a firm conclusion about the quasar variability owing to the controversial results of the light curves statistical analysis. In addition, we calibrated Johnson-Cousins $BVRI$ magnitudes of 7 field stars that are to be used as secondary standards. 
\end{abstract}

\section{INTRODUCTION}
Variability is a common property of active galactic nuclei (AGNs). It could be efficiently used to study their structure and properties. In particular, monitoring of high-luminosity quasars allows us to probe the near vicinity of most massive black holes \cite{Kaspi07}.

\hs\ is a high-redshift ($z$ = 3.051, \cite{Fan94}), high-luminosity ($M_{B}^{\rm abs}=-29.0\,\rm mag$, \cite{Veron10}), radio-quiet \cite{Bechtold94} quasar. It was discovered in the course of the Hamburg Quasar Survey \cite{Hagen92}. At that time this object was the most luminous quasar known. Its total luminosity is in excess of $5.7 \times 10^{48}\,\rm erg\,s^{-1}$ \cite{Hagen92}.

Since July 1997 we have initiated a photometric monitoring of \hs\ in order to do a characterization of its variability on various time scales: intra-night (from minutes to hours), short-term (from days to weeks), and long-term (from months to years).
The first results of our monitoring were presented in Ref. \cite{Mihov99}. No evidence for strong variability during one year of monitoring was reported. 

In this study we present the results from the optical and X-ray intra-night monitoring of \hs. To calibrate our photometry and facilitate further monitoring of the quasar, we establish a secondary standard sequence in its field. 

\begin{figure*}[t]
\begin{minipage}[t]{0.45\linewidth}
\centerline{\includegraphics[width=\linewidth]{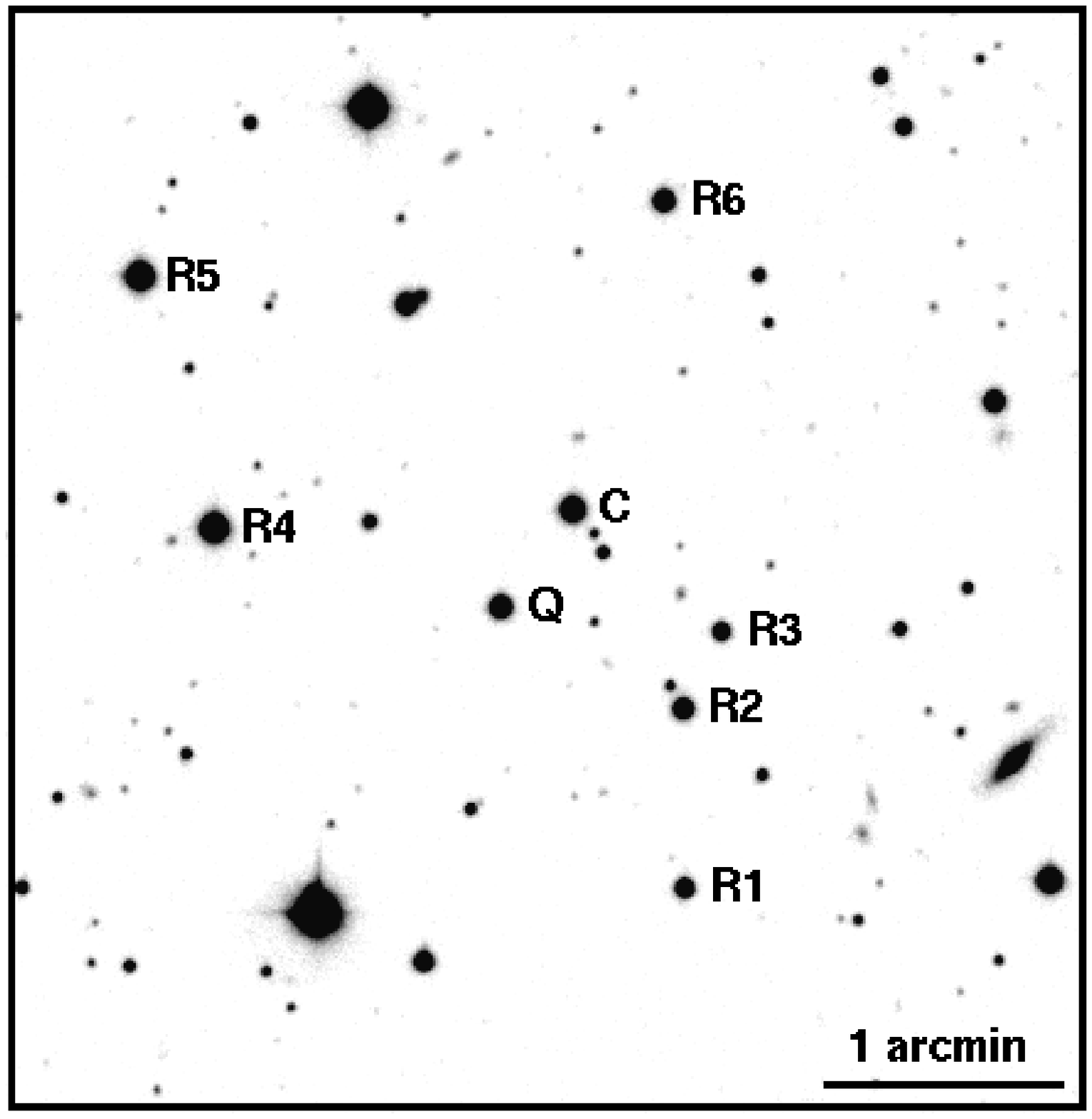}}
\caption{The field around the quasar \hs. The quasar, control star, and reference stars are labeled as Q, C, and R1 to R6, respectively. North is at the top, east to the left.}
\label{field}
\end{minipage}
 \hspace{0.6cm}
\begin{minipage}[t]{0.49\linewidth}
\centerline{\includegraphics[width=\linewidth]{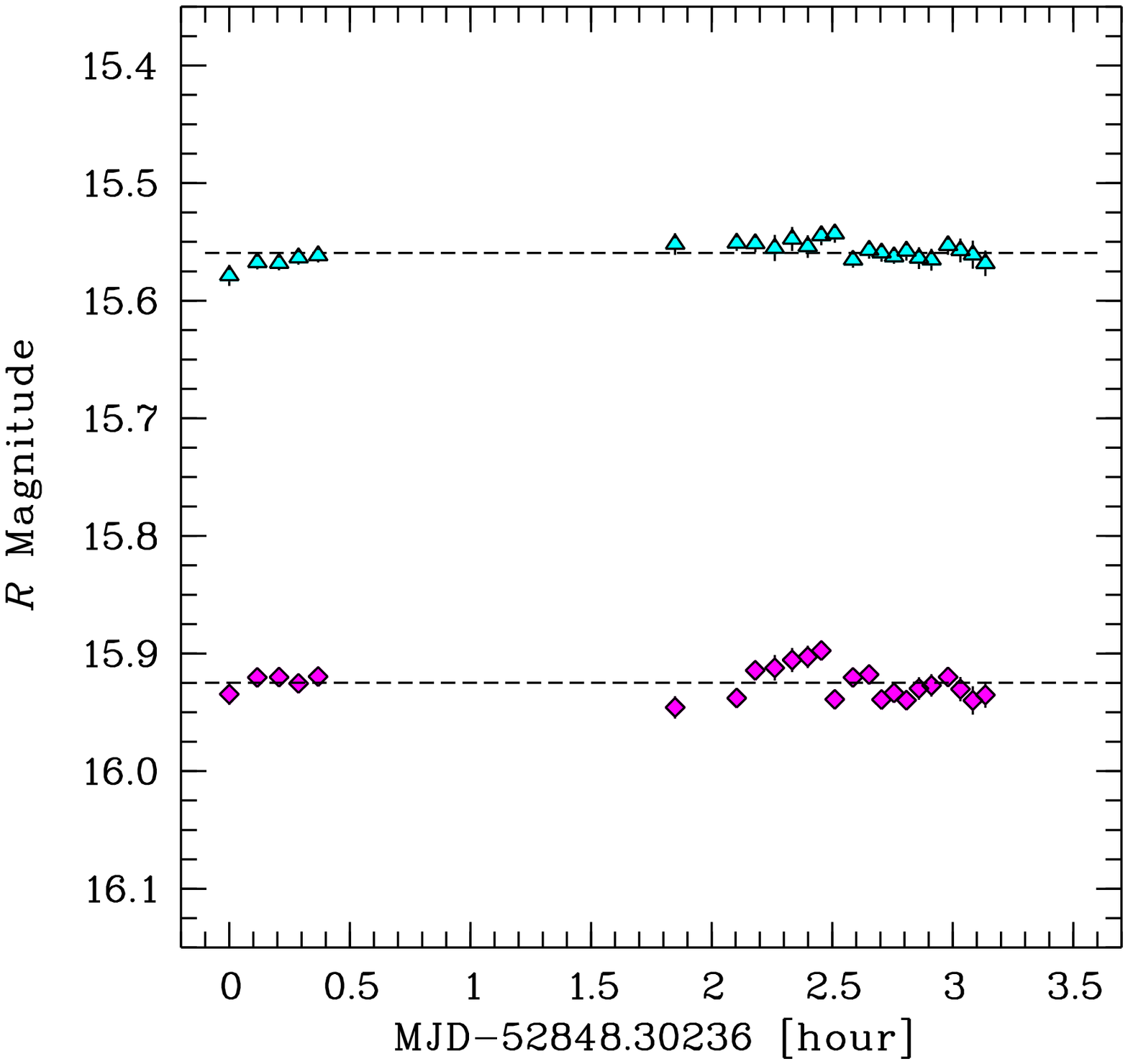}}
\caption{Intra-night light curve of the quasar (magenta diamonds) and of the control star (cyan triangles) for Jul 27, 2003. The dashed lines mark the corresponding weight-averaged magnitudes listed in Table\,\ref{inom}. The interruption is due to clouds.}
\label{at1}
\end{minipage}
\end{figure*}

\begin{table}[t]
\caption{Weight-averaged $BVRI$ magnitudes of the control and reference stars.}
\label{stds}
\tabcolsep7pt\begin{tabular}{l|cccc|cccc}
\hline
ID & $\langle m\rangle_{\rm w}$ & $\sigma_{\rm w}$ & $\chi^2_{\rm df}$ & $n$ & $\langle m\rangle_{\rm w}$ & $\sigma_{\rm w}$ & $\chi^2_{\rm df}$ & $n$ \\
\hline

   & $B$ & & & & $V$ & & & \\
\cline{2-2}
\cline{6-6}
\noalign{\smallskip}
C  & 16.713 $\pm$ 0.022 & 0.044 & 4.552 & 4 & 15.972 $\pm$ 0.015 & 0.034 & 3.180 & 5 \\
R1 & 17.895 $\pm$ 0.016 & 0.033 & 2.328 & 4 & 17.222 $\pm$ 0.014 & 0.031 & 2.591 & 5 \\
R2 & 17.975 $\pm$ 0.022 & 0.045 & 4.291 & 4 & 16.995 $\pm$ 0.018 & 0.041 & 4.487 & 5 \\
R3 & 18.465 $\pm$ 0.024 & 0.048 & 4.545 & 4 & 17.659 $\pm$ 0.017 & 0.039 & 4.029 & 5 \\
R4 & 16.270 $\pm$ 0.022 & 0.044 & 4.595 & 4 & 15.482 $\pm$ 0.010 & 0.021 & 1.250 & 5 \\
R5 & 16.459 $\pm$ 0.012 & 0.025 & 1.426 & 4 & 15.612 $\pm$ 0.011 & 0.024 & 1.632 & 5 \\
R6 & 18.829 $\pm$ 0.014 & 0.029 & 1.403 & 4 & 17.301 $\pm$ 0.012 & 0.027 & 1.892 & 5 \\
\hline                                                                                 

   & $R$ & & & & $I$ & & & \\                                                          
\cline{2-2}
\cline{6-6}
\noalign{\smallskip}
C  & 15.544 $\pm$ 0.011 & 0.011 & 0.209 & 5 & 15.144 $\pm$ 0.012 & 0.022 & 1.394 & 3 \\
R1 & 16.835 $\pm$ 0.011 & 0.020 & 0.661 & 5 & 16.426 $\pm$ 0.027 & 0.047 & 6.341 & 3 \\
R2 & 16.461 $\pm$ 0.011 & 0.018 & 0.528 & 5 & 15.971 $\pm$ 0.023 & 0.040 & 4.693 & 3 \\
R3 & 17.192 $\pm$ 0.011 & 0.017 & 0.457 & 5 & 16.728 $\pm$ 0.020 & 0.035 & 3.272 & 3 \\
R4 & 15.004 $\pm$ 0.011 & 0.014 & 0.330 & 5 & 14.561 $\pm$ 0.011 & 0.019 & 1.072 & 3 \\
R5 & 15.128 $\pm$ 0.011 & 0.022 & 0.851 & 5 & 14.651 $\pm$ 0.019 & 0.033 & 3.169 & 3 \\
R6 & 16.236 $\pm$ 0.014 & 0.030 & 1.861 & 5 & 15.085 $\pm$ 0.012 & 0.021 & 1.302 & 3 \\
\hline
\end{tabular}
\end{table}

\begin{figure*}[t]
\begin{minipage}[t]{0.47\linewidth}
\centerline{\includegraphics[width=\linewidth]{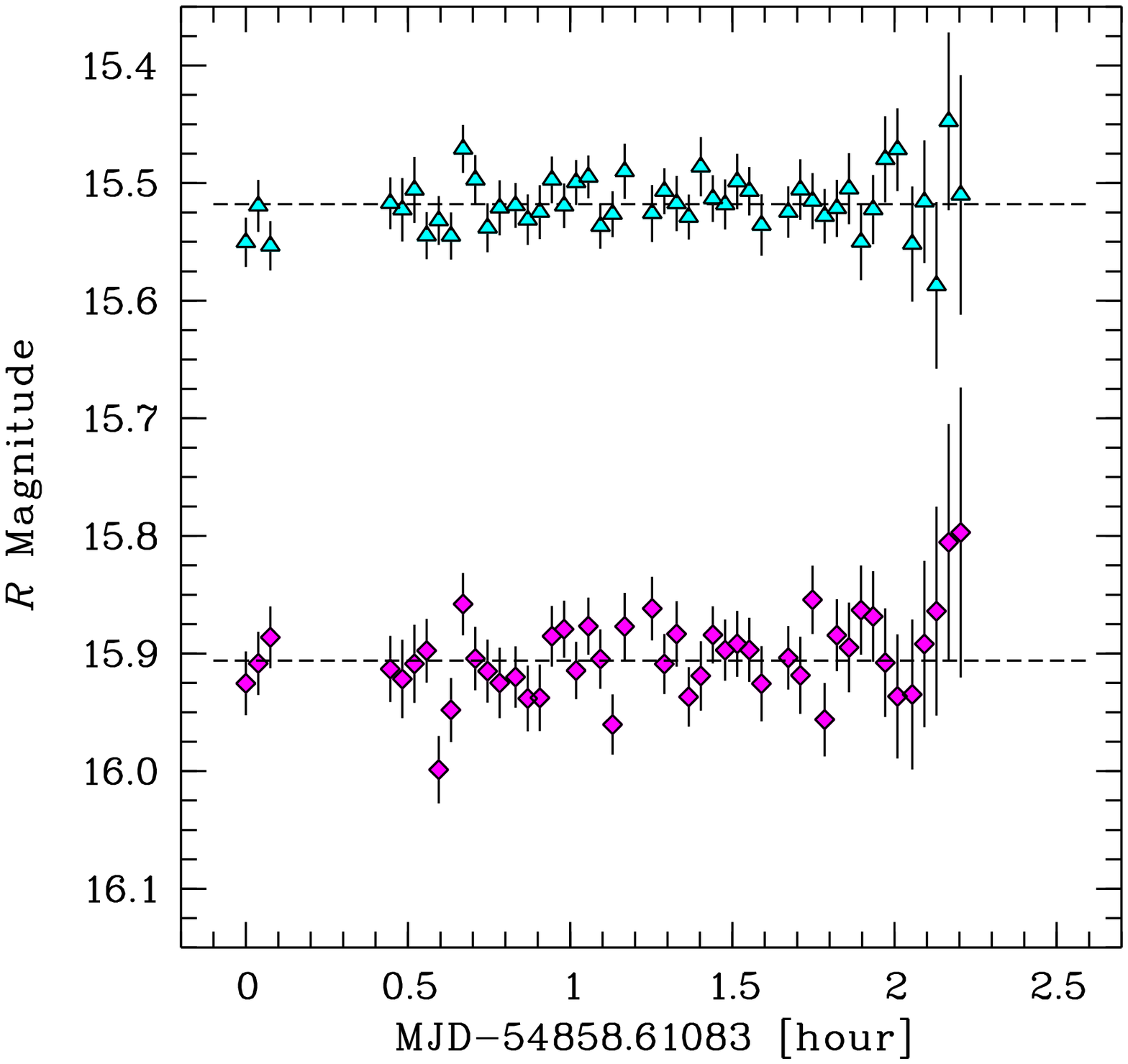}}
\caption{The same as in Figure\,\ref{at1}, but for Jan 26, 2009.}
\label{stl1}
\end{minipage}
 \hspace{0.6cm}
\begin{minipage}[t]{0.47\linewidth}
\centerline{\includegraphics[width=\linewidth]{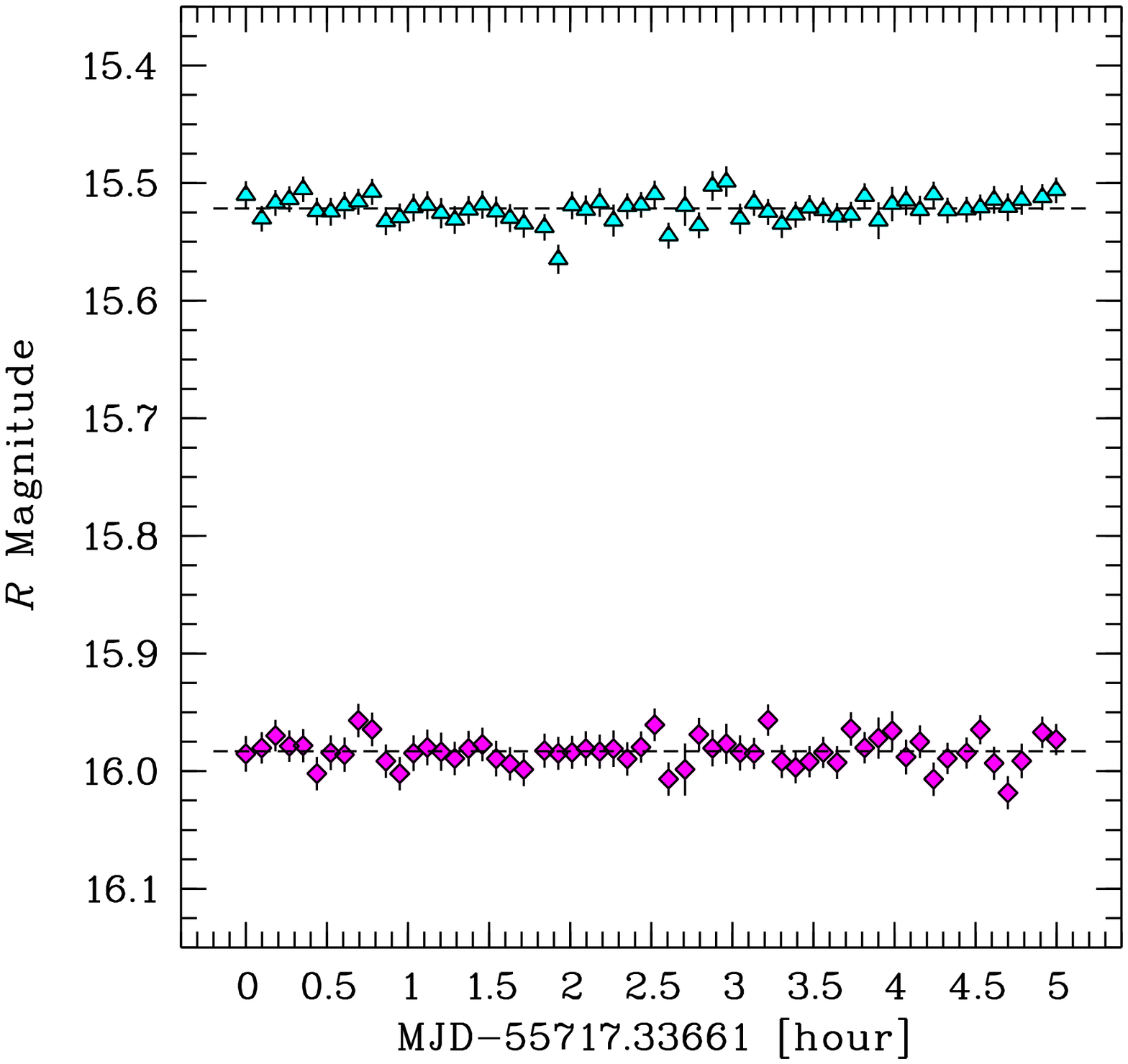}}
\caption{The same as in Figure\,\ref{at1}, but for Jun 4, 2011.}
\label{fli1}
\end{minipage}
\end{figure*}

\begin{table}[ht]
\caption{Intra-night monitoring results.}
\label{inom}
\tabcolsep7pt\begin{tabular}{ccccc}
\hline
Evening Date & Duration & $\langle R_{\rm QSO}\rangle_{\rm w}$ & $\langle R_{\rm control}\rangle_{\rm w}$ & $\kappa$ \\
$FWHM$ & $N$ & $\sigma_{\rm w,QSO}$; $\chi^2_{\rm df,QSO}$ & $\sigma_{\rm w,control}$; $\chi^2_{\rm df,control}$ & $C_{\kappa}$ \\
\hline
\noalign{\smallskip}
Jul 27, 2003 & 3.16 & $15.925 \pm 0.002$ & $15.559 \pm 0.002$ & 1.03 \\
1\farcs26; 0\farcs16  & 24   & 0.012; 2.229       & 0.008; 1.117       & 1.50 \\
\noalign{\smallskip}

Jan 26, 2009 & 2.17 & $15.906 \pm 0.004$ & $15.518 \pm 0.003$ & 1.25 \\
4\farcs04; 0\farcs58  & 48   & 0.030; 1.026       & 0.020; 0.732       & 1.18 \\
\noalign{\smallskip}
            
Jun 04, 2011 & 4.99 & $15.983 \pm 0.002$ & $15.522 \pm 0.002$ & 1.21 \\
2\farcs17; 0\farcs25  & 58   & 0.013; 0.788       & 0.011; 0.835       & 0.96 \\
\hline
\end{tabular}
\end{table}

\section{SECONDARY STANDARD SEQUENCE}

In order to calibrate secondary Johnson-Cousins $BVRI$ standard stars in the field of \hs, we observed the standard sequence established in the globular cluster M\,92 (e.g., \cite{Mihov17}) at least at two different airmasses. The observations were performed during the following nights:
\begin{itemize}
\item Jul 16, 2009, Aug 5, 6, 7, 2013, and Aug 20, 2014 in the $B$ band.
\item Jul 18,1998, Apr 17, 1999, Jul 16, 2009, Aug 5, 6, 7, 2013, and Aug 20, 2014 in the $VR$ bands.
\item Aug 5, 6, 7, 2013 and Aug 20, 2014 in the $I$ band.
\end{itemize}

We used the 2\,m telescope (2MT; \cite{Bonev10a}, \cite{Bonev10b}) of the Rozhen National Astronomical Observatory (NAO) equipped with Photometrics AT200\footnote{This camera has $1024\times1024$ pixels of $24\,\mu$m size resulting in a scale factor of 0\farcs309\,px$^{-1}$.} (1998 and 1999 data) and Princeton Instruments VersArray:1300B\footnote{This camera has $1340\times1300$ pixels of $20\,\mu$m size resulting in a scale factor of 0\farcs258\,px$^{-1}$.} (2009--2014 data) CCD cameras. The images of both \hs\ and M\,92 fields were de-biased, flat fielded (using twilight flats), de-fringed (regarding all $I$-band frames and the deepest $R$-band ones), cosmic ray hit cleaned, and co-added using {\sc eso-midas} and {\sc idl} packages. The photometry and transformation to the standard system were performed as in Ref. \cite{Mihov08}. We applied the spatial dependent systematic error correction only for the 1998 and 1999 data following Ref. \cite{Mihov17}.

The final weight-averaged magnitudes of the standard stars were estimated after removing up to two most deviant individual measurements. The finding chart of the standard sequence is shown in Figure\,\ref{field}. The star closest in position and magnitude to \hs\ (star C) is suitable to be used as a control star in the photometry of the quasar. The derived magnitudes of the stars are listed in Table\,\ref{stds}, together with the weighted standard deviation about the mean magnitude ($\sigma_{\rm w}$), the reduced figure-of-merit of the averaging ($\chi^2_{\rm df}$), and the number of the epochs being averaged ($n$). The standard stars show no intrinsic variability.

\section{INTRA-NIGHT OPTICAL MONITORING}

The Cousins $R$ band intra-night monitoring of \hs\ was performed during three nights with the 2MT and the 50/70\,cm Schmidt telescope (ST; \cite{Tsvetkov87}, \cite{Semkov97}, \cite{Kostov10}) of NAO with the following light detectors:
\begin{itemize}
\item AT200 CCD camera (at the 2MT) for Jul 27, 2003.
\item $4008\times2672$ SBIG STL-11000M CCD camera (at the ST, pixel size of 9\,$\mu$m or $1\farcs079$) for Jan 26, 2009.
\item $4096\times4096$ FLI PL16803 CCD camera (at the ST, pixel size of 9\,$\mu$m or $1\farcs079$) for Jun 4, 2011.
\end{itemize}

The exposure times of the quasar field were 150--300\,sec for the 2MT and 120--300\,sec for the ST\footnote{Due to tracking problems the exposure times at the ST are limited to few minutes.}.
All CCD frames were bias/dark subtracted and flat fielded using twilight flats. Aperture photometry of the objects of interest was then performed using {\sc daophot} package run under {\sc idl}\footnote{https://idlastro.gsfc.nasa.gov/}. The optimal aperture radii were set to $2\times FWHM$ arcsec in the case of 2MT, 5$^{\prime\prime}$ for ST/STL-11000M, and 4$^{\prime\prime}$ for ST/PL16803. 

For each frame we computed a zero-point with respect to each reference star, $zp_i=m_i-M_i,\,i=1,..,6$, where $m_i$ and $M_i$ are the instrumental
and standard magnitudes of the $i$-th reference star, respectively. Next, the zero-points were weight-averaged, $\langle zp \rangle_{\rm w}$. Finally, the most deviant zero-point was removed and the remaining ones were weight-averaged again. The calibrated magnitudes of the quasar and control star were obtained as $M=m-\langle zp \rangle_{\rm w}$. 
The resulting light curves are plotted in Figure\,\ref{at1} for the 2MT and in Figure\,\ref{stl1} and \ref{fli1} for the ST. The characteristics of the intra-night light curves are given in Table\,\ref{inom}. The first row lists the evening date, the monitoring duration in hours, the weight-averaged $R$ band magnitudes of the quasar and control star, and the scaling factor ($\kappa$, see below). The second row lists the median $FWHM$ during the monitoring, together with the standard deviation about it, the number of the light curve data points ($N$), the weighted standard deviation about the mean magnitudes ($\sigma_{\rm w}$), together with the reduced figure-of-merit of the averaging ($\chi^2_{\rm df}$), and the value of $C_\kappa$ (see below).
We did not take into account the colour coefficients, so, this resulted in differences in the magnitudes measured at the 2MT and ST. In addition, the quasar magnitudes are subject to its intrinsic long-term variability.

The search for intra-night variability was performed using the $C$ test \cite{Zibecchi17},
which compares the standard deviations of the quasar and the control star light curves:
$$
C=\frac{\sigma_{\rm QSO}}{\sigma_{\rm control}}.
$$
It states that a source is variable at 99.5\% confidence level if $C\ge2.576$.
Before applying the test we derived the scale factor:
$$
\kappa=\frac{\langle e_{\rm QSO}\rangle_{\rm med}}{\langle e_{\rm control}\rangle_{\rm med}},
$$
which is the ratio of the median uncertainty of the quasar and the control star magnitudes (e.g., \cite{Joshi11}).
Thus, the $C$ test is scaled as follows:
$$
C_{\kappa}=\frac{\sigma_{\rm QSO}}{\kappa\,\sigma_{\rm control}}.
$$
According to the values of $C_{\kappa}$, the quasar does not show variability at 99.5\% confidence level.
This result is consistent with the general lack of intra-night variability in high-luminosity, high-redshift radio-quiet quasars \cite{Bachev05}. 

\section{X-RAY LIGHT CURVES}

\begin{figure*}[t]
\begin{minipage}[t]{0.475\linewidth}
\centerline{\includegraphics[width=\linewidth]{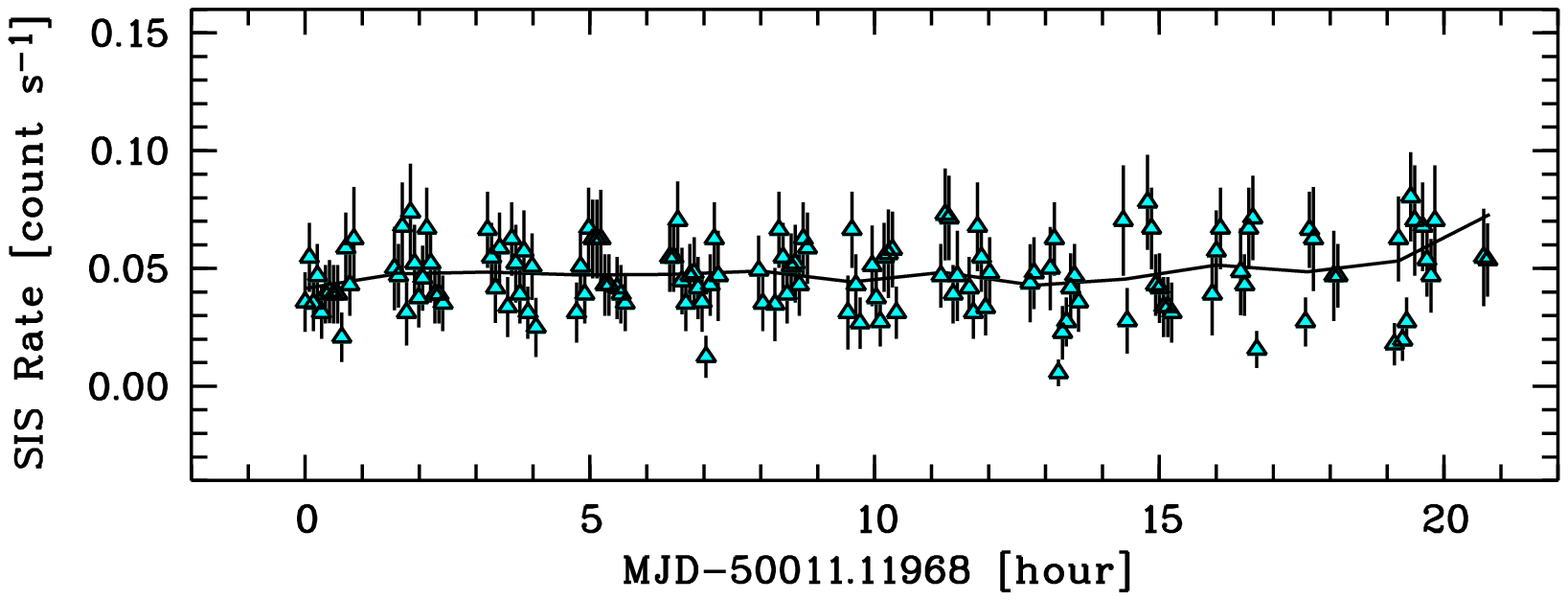}}
\caption{Combined SIS0+SIS1 0.5--10\,keV light curve with 256\,sec time resolution for Oct 21, 1995.
The solid line represents the same light curve with 5760\,sec time resolution, which is roughly the ASCA orbital period.}
\label{sis}
\end{minipage}
 \hspace{0.6cm}
\begin{minipage}[t]{0.465\linewidth}
\centerline{\includegraphics[width=\linewidth]{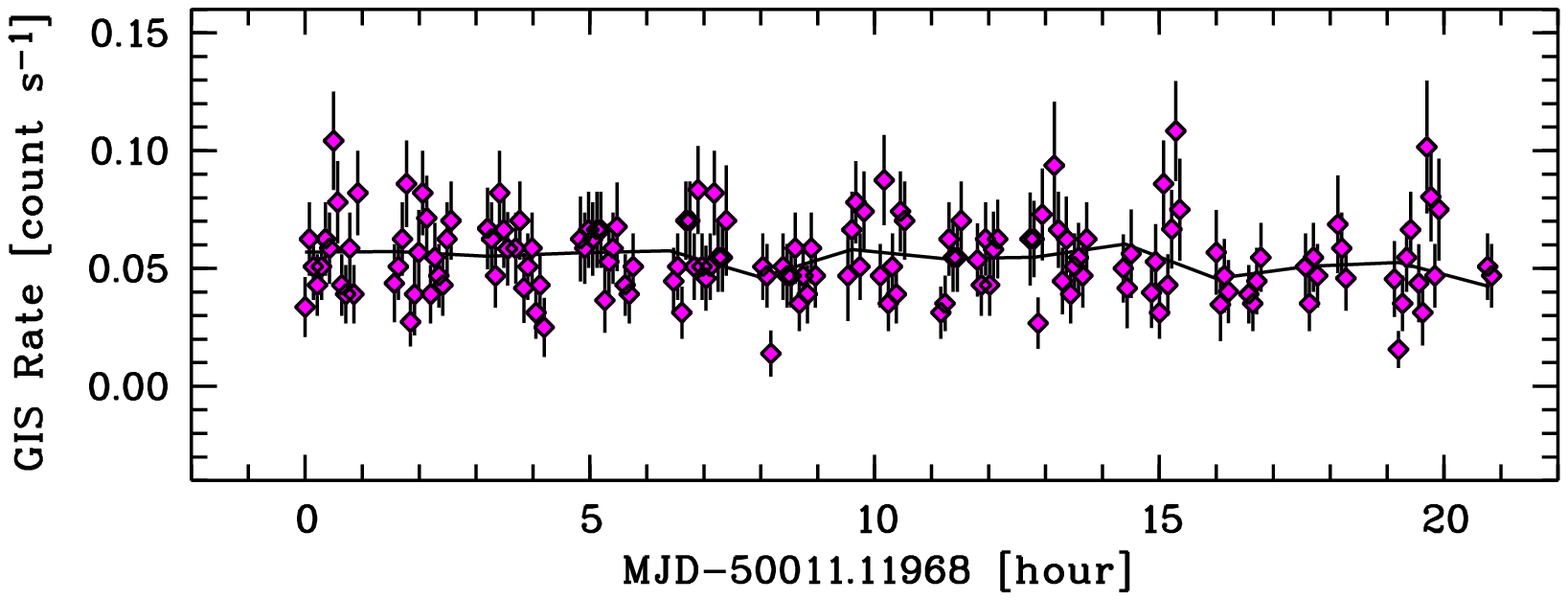}}
\caption{The same as in Figure\,\ref{sis}, but for the combined GIS3+GIS4 0.75--10\,keV light curve.}
\label{gis}
\end{minipage}
\end{figure*}

The X-ray light curves of \hs, obtained on Oct 21, 1995, were taken from the {\sc tartarus} database, which contains reduced images, spectra, and light curves for ASCA\footnote{Advanced Satellite for Cosmology and Astrophysics.} observations of AGNs \cite{Turner01}.
ASCA has four X-ray telescopes. Solid-state Imaging Spectrometers (SIS) are detectors for two of them, while Gas Imaging Spectrometers (GIS) are detectors for the other two. We show in Figure\,\ref{sis} and \ref{gis} the combined SIS0+SIS1 0.5--10\,keV and GIS2+GIS3 0.75--10\,keV light curves of \hs\ with 256\,sec time resolution, respectively.

The statistical $\chi^2$ test showed that, at 99.5\% confidence level, \hs\ is variable ($p$-value = 0.002) if we consider combined SIS light curves and non-variable ($p$-value = 0.07) if we consider combined GIS light curves. Owing to this controversial result, we cannot consider \hs\ variable in the X-rays given the ASCA data on the above date.

\section{SUMMARY}

Our results could be summarized as follows:
\begin{itemize}
\item We calibrated 7 stars to be used as secondary standards in the field of \hs.

\item We presented results of the $R$ band intra-night monitoring of \hs. The scaled $C$ test indicates no variability during almost 11 hours of monitoring. This is in accordance with the low intra-night variability duty cycle of radio-quiet quasars.

\item The statistical analysis of the X-ray light curves of about 22 hours duration shows controversial results regarding the quasar variability, so, we cannot make a firm conclusion about the quasar flux variation at that time.
\end{itemize}

\section{ACKNOWLEDGMENTS}
This research was partially supported by the Bulgarian National Science Fund of the Ministry of Education and Science under grants DN 08-1/2016 and DN 18/13-2017.
This research has made use of the {\sc tartarus} (Version 3.1) database, created by Paul O'Neill and Kirpal Nandra at Imperial College London, and Jane Turner at NASA/GSFC. {\sc tartarus} is supported by funding from PPARC, and NASA grants NAG5-7385 and NAG5-7067.

\end{document}